\documentclass[submission, Proceedings]{SciPost}

\hypersetup{
    colorlinks,
    linkcolor={red!50!black},
    citecolor={blue!50!black},
    urlcolor={blue!80!black}
}

\usepackage[bitstream-charter]{mathdesign}
\DeclareSymbolFont{usualmathcal}{OMS}{cmsy}{m}{n}
\DeclareSymbolFontAlphabet{\mathcal}{usualmathcal}

\newcommand{\ev}{\mbox{eV\,}$c^{-2}$}
\newcommand{\eve}{\mbox{eV$_{\rm ee}$}}
\newcommand{\um}{\mbox{$\mu$m}}

\begin{document}

\begin{center}{\Large \textbf{
The low-energy spectrum in DAMIC at SNOLAB\\
}}\end{center}

\begin{center}
Alvaro E. Chavarria\textsuperscript{1$\star$} for the DAMIC Collaboration
\end{center}

\begin{center}
{\bf 1} Center for Experimental Nuclear Physics and Astrophysics, University of Washington, Seattle, United States
\\
* chavarri@uw.edu
\end{center}

\begin{center}
\today
\end{center}


\definecolor{palegray}{gray}{0.95}
\begin{center}
\colorbox{palegray}{
  \begin{tabular}{rr}
  \begin{minipage}{0.1\textwidth}
    \includegraphics[width=30mm]{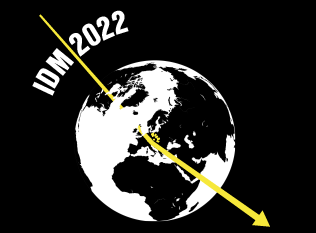}
  \end{minipage}
  &
  \begin{minipage}{0.85\textwidth}
    \begin{center}
    {\it 14th International Conference on Identification of Dark Matter}\\
    {\it Vienna, Austria, 18-22 July 2022} \\
    \doi{10.21468/SciPostPhysProc.?}\\
    \end{center}
  \end{minipage}
\end{tabular}
}
\end{center}

\section*{Abstract}
{\bf
The DAMIC experiment employs large-area, thick charge-coupled devices (CCDs) to search for the interactions of low-mass dark matter particles in the galactic halo with silicon atoms in the CCD target. From 2017 to 2019, DAMIC collected  data with a seven-CCD array (40-gram target) installed in the SNOLAB underground laboratory. We report dark-matter search results, including a conspicuous excess of events above the background model below 200\,\eve, whose origin remains unknown. We present details of the published spectral analysis, and update on the deployment of skipper CCDs to perform a more precise measurement by early 2023.
}

\section{Introduction}
\label{sec:intro}

The DAMIC experiment at SNOLAB employs the bulk silicon of scientific charge-coupled devices (CCDs) as a target for interactions of particle dark matter (DM) from the galactic halo.
The low pixel readout noise of 1.6\,$e^-$ R.M.S., combined with extremely low leakage current of a few $e^-$ per mm$^2\cdot$day, provides DAMIC CCDs with sensitivity to the small ionization signals from recoiling electrons or nuclei following the interactions of low-mass DM particles.

\section{DAMIC at SNOLAB}
\label{sec:detector}

DAMIC CCDs were developed by Berkeley Lab's Microsystems Laboratory and fabricated by Teledyne DALSA.
The devices feature a rectangular array of pixels, each of size 15$\times$15\,\um$^{2}$, and a fully-depleted active region of 675\,\um.
Other details of the CCD design and fabrication process can be found in Ref.~\cite{1185186}.
Several arrangements of CCDs were deployed in the DAMIC cryostat since 2012, with the final installation of seven 16\,Mpix CCDs in 2017, for a total silicon target mass of 40\,g.
Details of the DAMIC setup at SNOLAB are presented in Refs.~\cite{DAMIC:2016lrs, DAMIC:2021crr}.

\begin{figure}[t]
\centering
\includegraphics[width=0.9\textwidth]{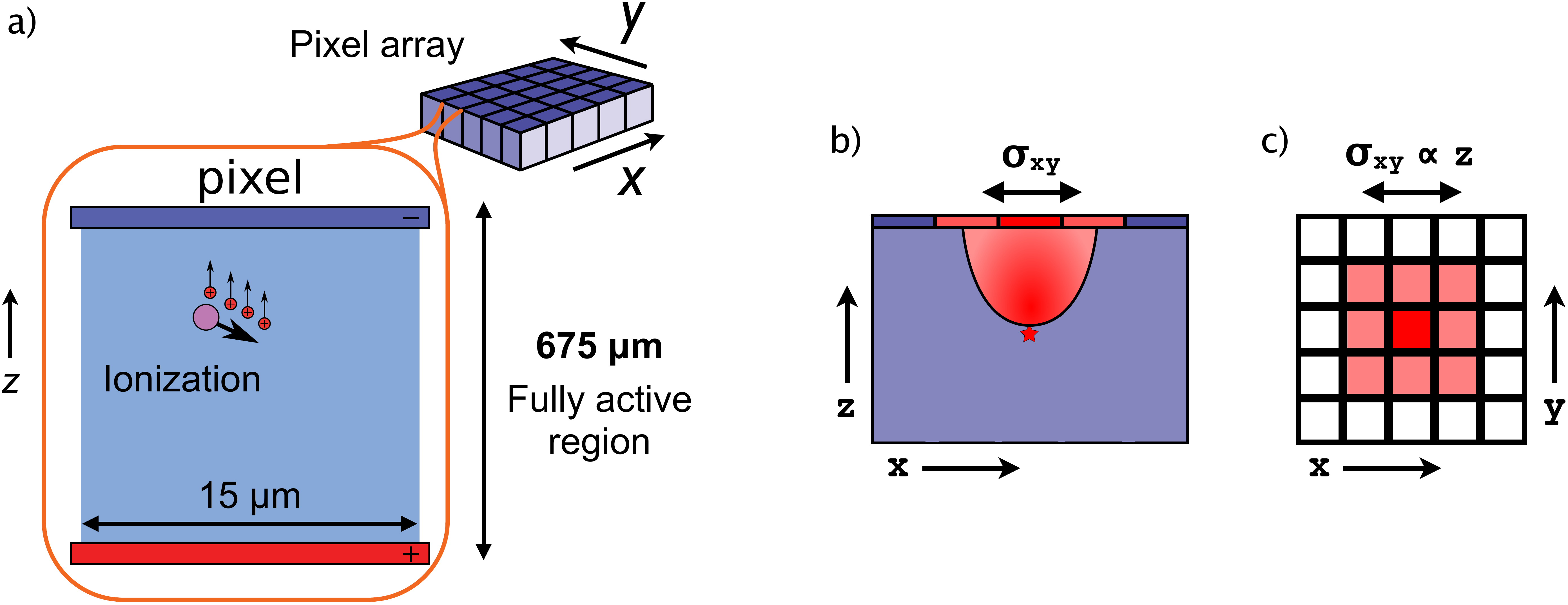}
\caption{{\bf a)}~Sketch of a low-energy particle ionizing the CCD active region. {\bf b)}~Diffusion of charge as it drifts toward the pixel array. {\bf c)}~The spread of the cluster of charge on the pixel array is positively correlated with the depth of the interaction.}
\label{fig:event}
\end{figure}

DAMIC data consists of images that contain a two-dimensional projection of all charge generated in the active region of the CCDs throughout an image exposure (typically 8\,h).
Particles generate free charges (e-h pairs) in the CCD active region by ionization (Fig.~\ref{fig:event}a), with one free e-h pair generated on average for every 3.8\,eV of kinetic energy deposited by a recoiling electron.
For a recoiling nucleus, the charge yield is lower and non-linear, and was directly calibrated in Ref.~\cite{Chavarria:2016xsi}.
The free charges are then drifted by an electric field toward the pixel array.
Since charge diffuses laterally as it drifts, interactions that occur deeper in the CCD bulk lead to more diffuse charge clusters (Fig.~\ref{fig:event}b).
The spread of the charge in the image ($x$-$y$ plane) can then be used to reconstruct the depth ($z$) of an interaction in the CCD active region (Fig.~\ref{fig:event}c).
Clustering algorithms are run on DAMIC images to identify clusters of pixels with charge above noise and reconstruct the deposited energy and $(x, y, z)$ location of particle interactions in the bulk silicon.
Details on the CCD response, image cleanup and processing are also presented in Refs.~\cite{DAMIC:2016lrs, DAMIC:2021crr}.

Dark matter searches in DAMIC are performed by comparing the charge (energy) distributions of individual pixels or pixel clusters against a background model that includes instrumental noise and ionizing backgrounds from natural radioactivity.
Ref.~\cite{DAMIC:2021crr} provides all details on the construction of the radioactive background model, including the extensive radioassay program of all components.
The background model was constrained and validated with several independent measurements of radiocontaminants in the detector, \emph{e.g.}, surface/bulk $^{210}$Pb and bulk $^{32}$Si, that were performed with the CCDs themselves by searching for spatio-temporal correlations between decays~\cite{DAMIC:2020wkw}.

The main science results from DAMIC are:
\begin{itemize}
\item The first search for DM interactions that produce as little as a one e-h pair in silicon, resulting in the first exclusion limit on the absorption of hidden photons~\cite{Hochberg:2016sqx} with masses as small as 1.2\,\ev~\cite{DAMIC:2016qck}.
\item Exclusion limits on the scattering of hidden-sector DM particles~\cite{Essig:2015cda} with masses as small as 0.5\,M\ev\ with electrons~\cite{DAMIC:2019dcn}.
\item The most sensitive direct search for weakly interacting massive particles (WIMPs)~\cite{Kolb:1990vq, Griest:2000kj, Zurek:2013wia} with masses in the range 1--9\,G\ev\ \cite{DAMIC:2020cut} scattering with silicon nuclei. This result significantly constrains any DM interpretation of the event excess observed by the CDMS-II Si experiment~\cite{CDMS:2013juh}, which employed the same nuclear target.
\end{itemize}

\section{Low-energy spectrum}
\label{sec:past}

\begin{figure}[t]
\centering
\includegraphics[width=\textwidth]{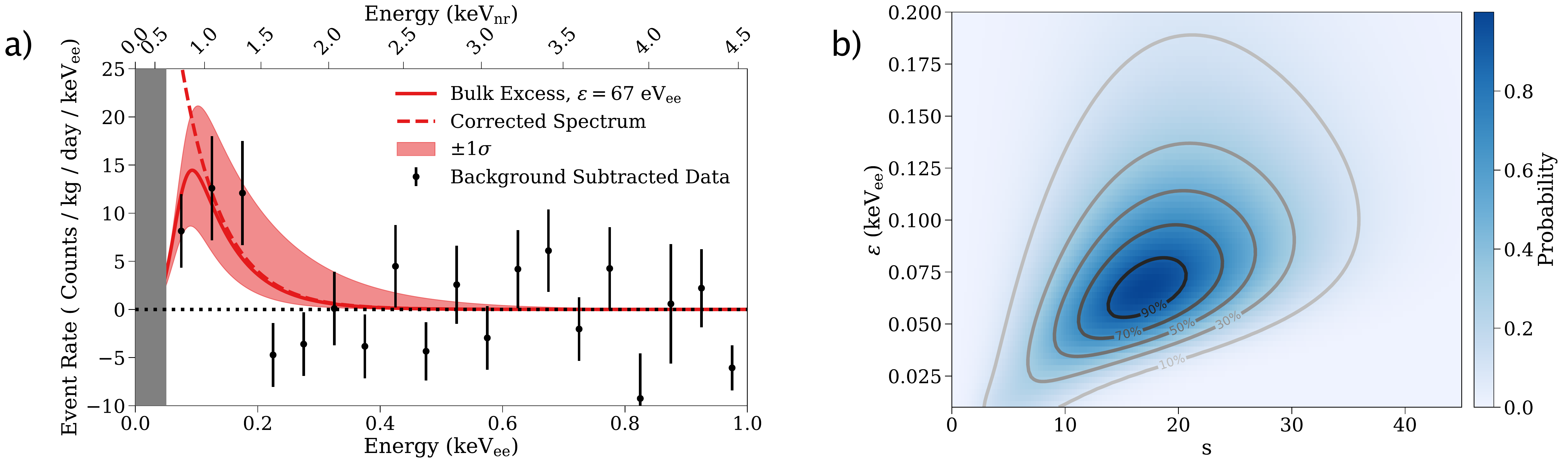}
\caption{{\bf a)}~Energy spectrum of the event excess (red lines) overlaid on the background-subtracted data (markers). Both the fit spectrum that includes the detector response (solid line) and the spectrum corrected for the detection efficiency (dashed line) are provided. The red shaded region represents the 1\,$\sigma$ uncertainty from the fit. {\bf b)}~Fit uncertainty in the number of excess events over the background model ($s$) and characteristic decay energy ($\epsilon$) of the exponential spectrum. The color axis represents the $p$-value from the fit. Figures from Ref.~\cite{DAMIC:2021crr}.}
\label{fig:excess}
\end{figure}

The background model developed for DAMIC's WIMP search describes the data remarkably well down to 200\,\eve .
Below this energy, and down to the detector's 50\,\eve\ threshold, there is a statistically significant excess (3.7\,$\sigma$) of events whose spatial distribution is consistent with being uniform in the bulk silicon and whose spectrum is well parametrized by a decaying exponential.
Figure~\ref{fig:excess}a shows the energy distribution of the excess of events above the background spectrum.
The $p$-value for the excess as a function of amplitude (number of events in the 11\,kg$\cdot$day exposure) and decay energy of the exponential is shown in Fig.~\ref{fig:excess}b.

Ref.~\cite{DAMIC:2021crr} presents a detailed investigation on the origin of the excess.
The events are significantly above threshold, and cannot arise from any known sources of instrumental noise (e.g., white readout noise or shot noise from leakage current).
Studies were performed on the spatial distribution of the events, and it was concluded that neither the known decaying spectrum from backside events that suffer from partial charge collection or an unknown population of frontside events could explain the excess.
Additionally, it was confirmed that besides the excess, there were no other statistically significant deviations from the background model throughout the full energy spectrum.
So far, the event excess is consistent with an unknown source of ionization events in the bulk silicon with a rate of a few per kg$\cdot$day.

\section{Skipper-CCD Upgrade}
\label{sec:another}
\begin{figure}[t]
\centering
\includegraphics[width=\textwidth]{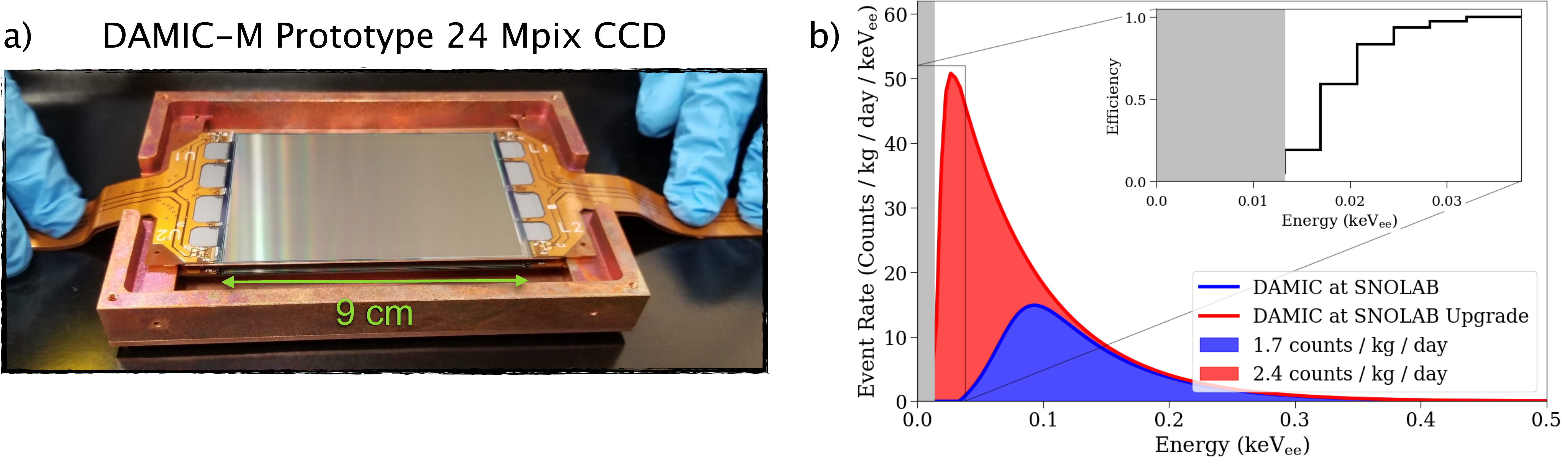}
\caption{{\bf a)}~DAMIC-M 24\,Mpix CCDs in their copper storage box. The bottom CCD is barely visible under the top CCD. {\bf b)}~Exponentially decaying spectrum of the excess of events observed in DAMIC at SNOLAB. The blue spectrum is the best-fit from the published analysis~\cite{DAMIC:2020cut, DAMIC:2021crr}, while the red spectrum is the expectation for the upgraded detector with skipper CCDs. The measured detection efficiency of the upgraded detector (inset) was used to construct the red spectrum.}
\label{fig:upgrade}
\end{figure}

To confirm the presence of the excess, and to better understand its origin, novel skipper CCDs were deployed in the DAMIC cryostat at SNOLAB.
This activity is carried out in collaboration with the DAMIC-M and SENSEI Collaborations, who provided the skipper CCDs and readout electronics~\cite{Cancelo:2020egx}.
Figure~\ref{fig:upgrade}a shows the two 24\,Mpix DAMIC-M prototype skipper CCDs being installed in their copper storage box during deployment, which amount to a silicon target of 18\,g.
The detector upgrade was completed in November 2021 and, following a commissioning period, science-data taking started in March 2022.
So far, 3\,kg$\cdot$day of data have been acquired with a total (bulk) radioactive background rate of 9 (6) events per k\eve $\cdot$kg$\cdot$day, comparable to the previous detector installation.
Data taking will continue throughout 2022 with first results expected in early 2023.

Skipper CCDs feature pixel readout noise as low as 0.05\,$e^-$, with the capability of counting with high resolution the number of charges collected in a pixel~\cite{Tiffenberg:2017aac}.
The much lower noise will decrease the threshold from 50\,\eve\ to 15\,\eve , and significantly improve the measurement of the $x$-$y$ spread of the clusters for better $z$ resolution.
Figure~\ref{fig:upgrade}b shows the much larger fraction of the exponentially decaying spectrum that is expected to be observed with skipper CCDs.
The skipper-CCD spectrum assumes the noise performance of the detector measured at SNOLAB after deployment, with the detection efficiency provided in the inset.

\section{Conclusion}
Over the past decade, DAMIC pioneered the use of low-noise CCDs in a low-radioactivity underground environment to search for DM.
The seven-CCD setup of DAMIC at SNOLAB that operated from 2017 to 2019 accrued the largest exposure with the lowest radioactive background of any CCD DM detector to date.
DAMIC published results from several DM searches, placing stringent exclusion limits on the interaction cross sections of DM particles with both electrons and nuclei in silicon atoms.
In the latest search for WIMPs, a conspicuous excess of events was observed above the background model below 200\,\eve, whose origin remains unknown.
The DAMIC detector at SNOLAB was recently upgraded with ultra-low noise skipper CCDs to clarify the origin of these events.

\section*{Acknowledgements}
These proceedings report on the work of the DAMIC Collaboration, currently composed of 38 scientists from 10 institutions around the world.
The DAMIC experiment is made possible by SNOLAB and its staff, who provide underground space, logistical and technical services.

\bibliography{myrefs.bib}

\nolinenumbers

\end{document}